\pdfoutput=1
\documentclass[aps,prl,preprint,superscriptaddress,footinbib]{revtex4-2}
\usepackage{graphicx}

\begin{document}


\title{Coercivity Mechanisms of Single-Molecule Magnets}


\author{Lei Gu}
\email[]{gulei@sicnu.edu.cn}
\affiliation{College of Physics and Electronic Engineering, Sichuan Normal University, Chengdu 610101, China}
\author{Qiancheng Luo}
\affiliation{Frontier Institute of Science and Technology (FIST), Xi’an Jiaotong University, Xi’an 710054, Shaanxi, China}
\author{Guoping Zhao}
\email[]{zhaogp@uestc.edu.cn}
\affiliation{College of Physics and Electronic Engineering, Sichuan Normal University, Chengdu 610101, China}
\author{Yan-Zhen Zheng}
\email[]{zheng.yanzhen@xjtu.edu.cn}
\affiliation{Frontier Institute of Science and Technology (FIST), Xi’an Jiaotong University, Xi’an 710054, Shaanxi, China}
\author{Ruqian Wu}
\email[]{wur@uci.edu}
\affiliation{Department of Physics and Astronomy, University of California, Irvine, California 92697, USA}



\begin{abstract}
Magnetic hysteresis has become a crucial aspect for characterizing single-molecule magnets, but the comprehension of the coercivity mechanism is still a challenge. By using analytical derivation and quantum dynamical simulations, we reveal fundamental rules that govern magnetic relaxation of single molecule magnets under the influence of external magnetic fields, which in turn dictates the hysteresis behavior. Specifically, we find that energy level crossing induced by magnetic fields can drastically increase the relaxation rate and set a coercivity limit. The activation of optical-phonon-mediated quantum tunneling accelerates the relaxation and largely determines the coercivity. Intra-molecular exchange interaction in multi-ion compounds may enhance the coercivity by suppressing key relaxation processes. Unpaired bonding electrons in mixed-valence complexes bear a pre-spin-flip process, which may facilitate magnetization reversal. Underlying these properties are magnetic relaxation processes modulated by the interplay of magnetic fields, phonon spectrum and spin state configuration, which also proposes a fresh perspective for the nearly centurial coercive paradox.
\end{abstract}


\maketitle


\paragraph{Introduction.} The field of magnetism continues to grapple with long-standing topics of coercivity mechanisms, especially the paradox of notably lower coercivity compared to the theoretical predictions~\cite{Brown1945,Brown1963,Aharoni1960,Aharoni1996,Hartmann1987,Brandon1992,Bertotti1998,Zhao2006,Zhao2019}. This discrepancy is usually referred to as Brown's coercive paradox since it was brought up in the 1940s~\cite{Brown1945}. While diverse mechanisms such as imperfection~\cite{Aharoni1960}, inter-grain interactions~\cite{Aharoni1996}, boundary effects~\cite{Hartmann1987}, and nonlocal exchange interaction~\cite{Brandon1992} have been proposed to bridge the gap between  theory and experiment, the paradox has not been satisfactorily resolved, and the underlying physics is still a subject of ongoing debate. The paradox is especially prevalent when considering the magnetic hysteresis measurements of single-molecule magnets (SMMs)~\cite{Goodwin2017,Guo2018,Randall2018,Gould2022,Rinehart2011a,Rinehart2011b,Guo2011,Demir2012,Langley2013,Chen2016,Demir2017,Ding2018,Liu2017,Yu2020,Huang2016,Santana2022,Andoni2021,Ding2022,Ling2022}. The coercivity mechanism in SMMs essentially links to how magnetic fields impact the relaxation processes of individual spins, which manifests as non-trivial imprints in the magnetic hysteresis behavior. As demagnetization at finite temperature and magnetic fields can be generally considered as magnetic relaxation~\cite{Nishino2020,Nishino2023}, investigation into these magneto-modulation effects in SMMs could also illuminate coercivity mechanisms in other magnetic materials.

SMMs have attracted great interests in recent decades for potential applications in quantum information technologies~\cite{Gaita2019,Coronado2020,Chilton2022}. Many efforts have been paid to the slow magnetic relaxation of SMM~\cite{Harman2010,Vallejo2012,Zhu2013,Zadrozny2013,Lucaccini2014,Gomez2014,Rajnak2019,JinWang2019,Kobayashi2019,Cui2019,Vallejo2019,Gu2020}, as 
the ability to maintain magnetization underpins their functionality. Recently, high temperature magnetic hysteresis in dysprosocenium~\cite{Goodwin2017,Guo2018,Randall2018} and ultrahard magnetism in mixed-valence dilanthanide complexes ~\cite{Gould2022} have marked advancements toward practical applications. These developments necessitate an in-depth investigation into magnetic relaxation of SMMs under the influence of strong magnetic fields. Despite various mechanisms of the magnetic relaxation of SMMs being available~\cite{Gomez2014, Ding2018,Shrivastava1983,Giraud2001,Ishikawa2005,ChenYC2017}, a theoretical description of magnetization evolution in strong magnetic fields has not been established. Fundamentally, it is still unclear what the coercivity of SMMs actually entails.

As the inter-molecular exchange interactions are negligible, SMMs are paramagnetic molecular crystals of magnetic complexes. Their magnetization either aligns to an external magnetic field or decays in the absence of a field. As a result, magnetic hysteresis becomes obvious only when the magnetic relaxation is slow compared to the sweep rate of the scanning magnetic field. Especially, the width of hysteresis curves depends on the rate of magnetization reversal under an opposing magnetic field. While specific shape of a hysteresis loop may varies with the experimental setup, understanding coercivity of SMMs essentially involves elucidating how the magnetization evolves under different magnetic fields.

In classical approaches, the magnetization of a system at zero temperature can be determined by minimizing the energy, with the thermal effects incorporated via empirical thermal activation~\cite{Bertotti1998} and stochastic dynamics~\cite{Palacios1998,Nishino2015}. The coercivity can be then derived from behavior of the magnetization in a magnetic field. Given that the magnetization of each molecule in a SMM evolves rather independently, its coercivity at zero temperature can be studied through a simple classical model for a single molecule. However, as magnetic states of SMMs are quantized and the transition rates among spin states depends on their wavefunctions, a classical model is unsuitable for describing their hysteresis behavior. In this study, we carry out quantum dynamical simulation of magnetic relaxation processes under generic settings. Our results account for a broad range of experimental observations, and the effects of more specific features can be deduced based on the mechanisms that are elucidated.

\paragraph{Critical strengths of external magnetic fields.} In both model analysis and numerical simulation conducted in this work, we set the initial states to the saturation magnetization along the $-z$ direction, i.e., classical magnetic moments $S_z=-S$  or the quantum counterpart $|S_z=-S\rangle$, where $S$ denotes the magnitude of the moments. Instead of plotting the hysteresis curve, which depends on sweep rate of the scanning field, we address the more general problem of how magnetization evolves under an external magnetic field, particularly one in the opposite direction ($+z$). As an example, we set $S=5/2$ in this work. However, the results can readily be adapted to other scenarios.

We first consider the single-ion compounds, in which each magnetic molecule contains a single metallic atom wrapped within organic groups. The magnetic moment, characterized by an uniaxial anisotropy, is described by the Hamiltonian $\mathbf{H}_s = -DS_{z}^{2}$. According to the classical approach, a magnetic field is needed to overcome the anisotropy barrier between the two ground states ($S_z=\pm S$) to cause a flip in the magnetization. Once the critical value is reached, an abrupt reversal in magnetization occurs due to the elevated magnetic potential energy. The coercivity is given by $H_c=2DS/\mu_B g$~\footnote{See Supplementary Material, which also includes Refs.~\cite{Handzlik2020,Magott2022,Brzozowska2021,Gu2022,Lunghi2022,Briganti2021,Reta2021,Kragskow2023,Gatteschi2011,Landau1991,Abragam2012,Chibotaru2012} and gives the details of the classical model analyses and quantum dynamical simulations, as well as animations of the magnetic relaxation processes.} with $\mu_B$ the Bohr magneton and g the g-factor. Assuming moderate values $S=5/2$ and $D=2$ meV, the coercive field is $ H_c=82.3$ T, which is an order higher than what is typically observed experimentally. It is apparent that there must be acceleration mechanism at play to facilitate fast magnetic relaxation at lower magnetic field, beyond the standard understanding for the swift magnetization reversal around the coercive field.

\begin{figure}
	\includegraphics[width=0.6\textwidth]{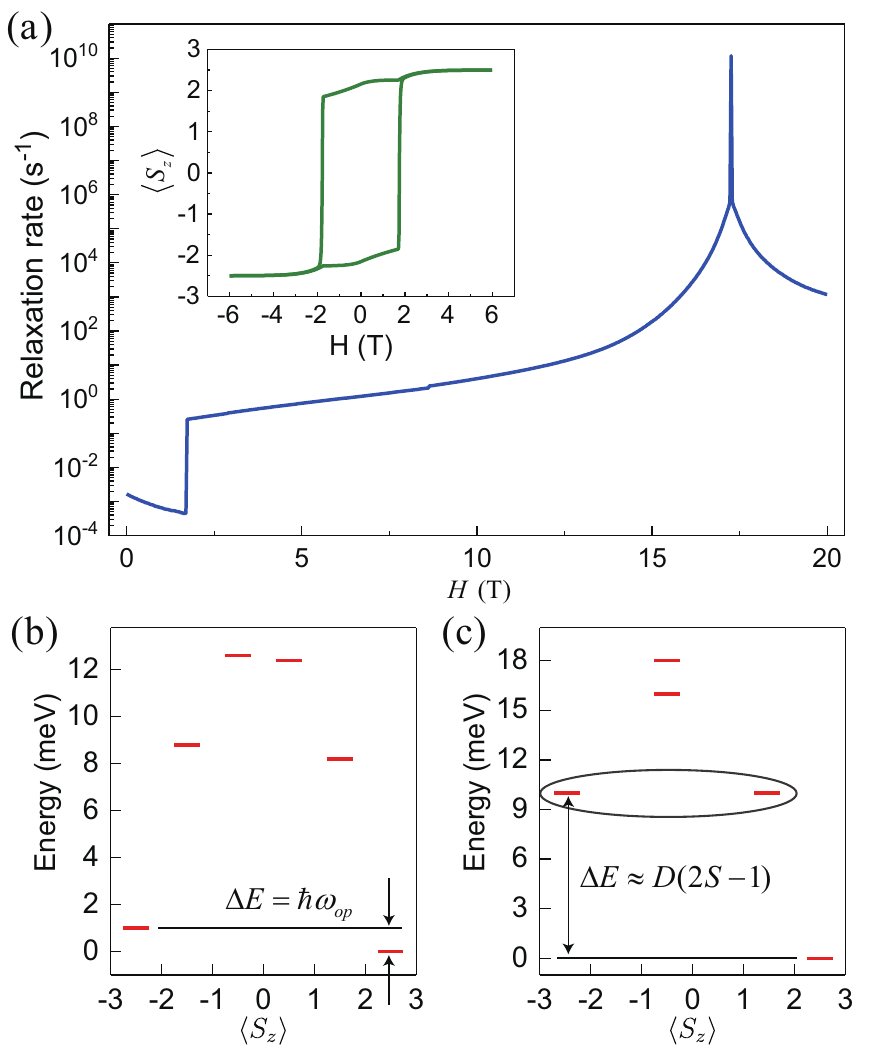}
	\caption{\label{Fig01} (a) The relaxation rate varies by more than ten orders under different magnetic fields. The fast magnetic reversal in the hysteresis loop is due to the opening up of the optical-phonon-mediated direct process between the ground states when the energy splitting matches energy of the significant optical phonons, as sketched in (b). (c) When energies of the circled states get close, the relaxation rate skyrockets and reaches the maximum at the level crossing point, corresponding to the peak region in (a).}
\end{figure}

Our computation suggests that the magnetic field can induce variation of relaxation rate by more than ten orders at a fixed temperature (such as $5$ K in this case), suggesting substantial magneto-modulation effects. We find that the rate maximum in Fig.~\ref{Fig01}(a) is related to level crossing induced by the magnetic field [Fig.~\ref{Fig01}(c)]. In our setting the magnetic field points to the $+z$ direction, so the energies of states with $|S_z<0\rangle$ increase with magnetic field, and the energies of states with $|S_z>0\rangle$ decrease.  The peak region of Fig.~\ref{Fig01}(a) suggests that the relaxation rate skyrockets when the energies of $|S_z= -S\rangle$ and $|S_z= S-1\rangle$ get close and peaks at the level crossing point. The soaring relaxation rate implies that the magnetization would reverse at an enormous speed. Therefore, it sets an effective limit on the coercivity, which approximates to $H'_c\approx D/\mu_B g$.

The skyrocketing increase of the relaxation rate can be interpreted using quantum perturbation theory. In addition to the uniaxial magnetic anisotropy, SMMs also exhibit slight transverse magnetic anisotropy, which at the lowest order takes the form $E(S_x^2-S_y^2)$. Since the transverse anisotropy does not commute with $S_z$, it leads to mixture of eigenstates of $S_z$ i.e., $|S_z=m\rangle$ with $m$ different integer values in the range $[-S, S]$. For $E\ll D$, the state mixing is perturbative, and an eigenstate of the combined Hamiltonian $\mathbf{H}_s=-D S_z^2+E(S_x^2-S_y^2)$ consists of predominantly one of the $|S_z=m\rangle$ and small proportions of $|S_z=n\rangle$ with $n\neq m$. The mixing weights initially scale as $(E/D)^{|m-n|}$ and increase when the energy difference $|E_m-E_n|$ is reduced by magnetic fields. In particular, two states may mix equally by a $50-50$ proportion, if their energies degenerate, regardless of the weakness of the transverse anisotropy. The most influential is the mixing of $|S_z= S-1\rangle$ into $|S_z= -S\rangle$, since it bridges direct tunneling between the ground states.

Taking the rotational spin-phonon coupling~\cite{Leuenberger2000,Calero2006,Ho2018,Chiesa2020} that reads $\mathbf{H}_{sp}\propto\sum_{\alpha=x,y,z}[\mathbf{H}_s, S_{\alpha}]$ for instance, the action of $\mathbf{H}_s$ does not cause overlap among the eigenstates of itself (i.e., spin eigenstates). It is the action of $S_x$ and $S_y$ that contributes to a considerable transition product $\langle s_2|\mathbf{H}_{sp}|s_1\rangle$, where $|s_1\rangle$ and $|s_2\rangle$ are two spin eigenstates. As $\mathbf{H}_{sp}$ is first order in $S_x$ and $S_y$, it imbricates two states $|S_z=m\rangle$ and $|S_z=n\rangle$ only when $|m-n|\leq 1$. The mixing weight of $|S_z= S-1\rangle$ in the elevated ground state built on $|S_z= -S\rangle$ is decisive for the transition between the two ground states, because in this case the weight times the term $\langle S_z=S-1|\mathbf{H}_{sp}|S_z=S\rangle$ approximates to the transition product. Other forms of spin-phonon coupling may cause state overlap with a broader span, i.e., $|m-n|\leq\Delta S_z$ with $\Delta S_z>1$. The rate skyrocketing persists, since $\langle S_z=S-1|\mathbf{H}_{sp}|S_z=S\rangle$ is still a major component of the transition product.

Being a result grounded in sound theory, the skyrocketing relaxation rate should lead to noticeable experimental signatures. Their absence in previous observations may be attributed to the use of scanning fields, where the demagnetization at lower field obscures the effect. Moreover, since the rate skyrocketing is triggered by the same condition with resonant tunneling~\cite{Gatteschi2003}, it might partially contribute to previous observations of resonant tunneling of multi-ion SMMs~\cite{Barbara1998,AKent_2000,Zhong2000,Chiorescu2000}, especially when the temperature is not close to zero. We should point out that the rate skyrocketing has nothing to do with the resonant tunneling between the degenerate states~\cite{Gatteschi2003}. Rather, it characterizes fast phonon-mediated transition between the ground states (one is lifted), and the excited state built on $|S-1\rangle$ itself is not an influential intermediate states.

However, we note that $H'_c=D/\mu_B g$ is still noticeably too high compared to what is typically observed in experiments. For example, $H'_c=17.3$ T when $D=2$~meV. Alternatively, the relaxation rate leap around $1.7$ T is consistent with typical observations. We find that this critical value corresponds to activation of the direct process between the ground states mediated by optical phonons. To open up this relaxation channel, the energy difference between the two states should match the energy of optical phonons [Fig.~\ref{Fig01}(b)]. As the Bose-Einstein distribution implies that the availability of a phonon decays exponentially as its energy increases, the low energy optical phonons with effective spin-phonon coupling dominate the process. The energy match implies $\mu_B g H\left[S-(-S)\right] \approx \hbar\omega_{op}$, where $\omega_{op}$ denotes frequency of the significant optical phonons. Thus, the critical field strength is given by $H''_c\approx\hbar\omega_{op}/2S\mu_B g$. $H''_c\approx 1.7$ T results from our setting $\omega_{op}=1$ meV and $S=5/2$.

To more comprehensively explain our proposed coercive mechanism, let us revisit an opinion advocated in previous work~\cite{Gu2020,Gu2021,Briganti2021,Ma2022}: the optical phonons dominate the magnetic relaxation of SMMs, while the acoustic phonons play a minor role~\footnote{In Sec. II A of the Supplementary Materials we provide more arguments and supportive simulation results for this fact.}. This is ascribed to the weak coupling between acoustic phonons and the magnetic state of a molecule, which in turn is because the molecules in SMMs are mechanically rather independent, due to the strong intra-molecular and weak inter-molecular interactions. On the other hand, the Raman processes are second order effects and in general weaker than direct processes. Therefore, once activated by relatively strong magnetic field, the optical-phonon-meditate direct transition between the ground states shall dominate the magnetic relaxation and largely determine the coercivity. In other words, the coercivity of SMMs is a manifestation of this magneto-phononic modulation effect.

\begin{figure}
	\includegraphics[width=0.6\textwidth]{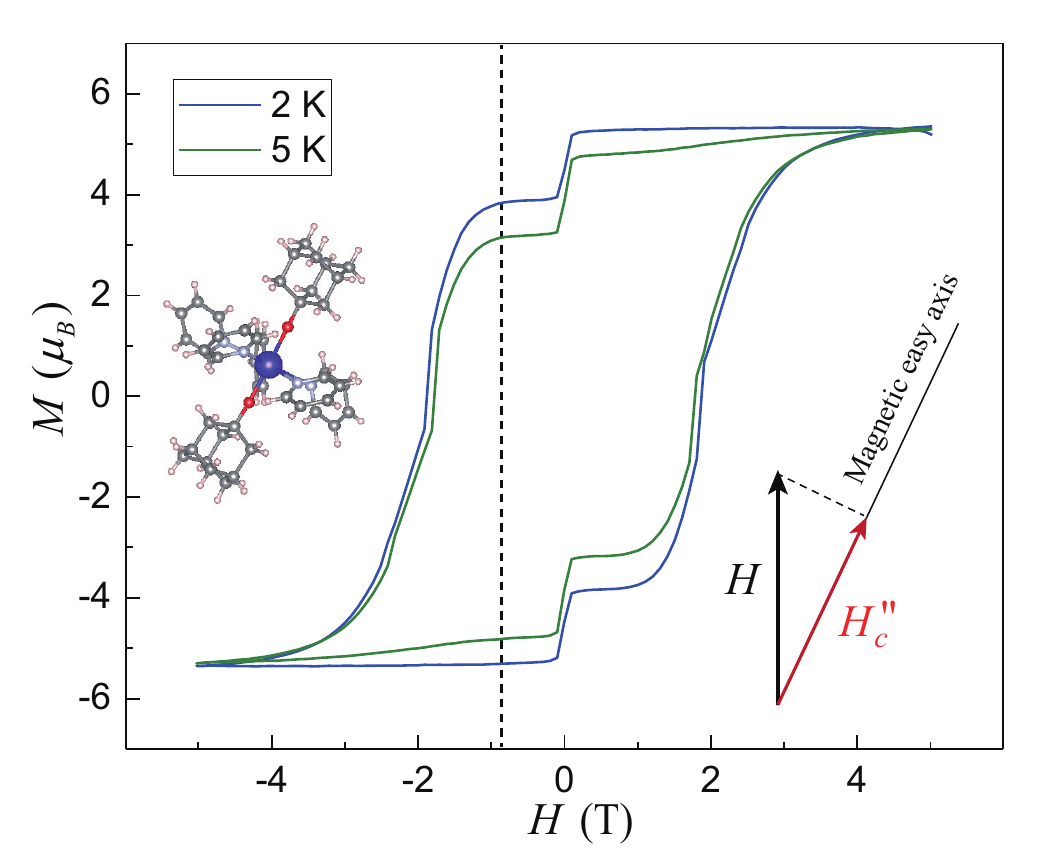}
	\caption{\label{Fig02} Measured magnetic hysteresis loops for the [Dy(1-AdO)$_2$(py)$_5$ ]BPh$_4$ complex. As the g-factor is completely biased to the magnetic easy axis, the projection of the magnetic field onto the easy axis is actually responsible for the Zeeman effect. When the magnitude of the projection reaches $H_c''$, rapid magnetic relaxation of the corresponding molecule is activated. In a polycrystal the activation starts at $H=H_c''$ (marked by the dashed line) and proceeds to involve molecules whose easy axes do not align with $H$.}
\end{figure}

\paragraph{Experimental evidences.} In our derivation and simulations, we assume a magnetic field aligning with the easy axis. For polycrystals of compounds with an anisotropic g-factor, the activation of the rapid reversal proceeds gradually and includes more molecules as the field strength increases. For instance, dysprosium complexes usually have a g-factor biased to the magnetic easy axis, and the two transverse components are close to zero. In this case, $H_c''$ should correspond to the projection onto the magnetic easy axis, as sketched in the right inset of Fig.~\ref{Fig02}. The larger is the angle between the easy axis and the magnetic filed, the field should be stronger to open the gap ($\Delta E$) in Fig.~\ref{Fig01}(b). Therefore, in polycrystalline samples, $H_c''$ is the critical field strength at which the acceleration of magnetization reversal starts, and the g-factor anisotropy smoothes the overall reversal and enhances the coercivity, which can be seen by comparing the inset of Fig.~\ref{Fig01}(a) and the experimental hysteresis loop in Fig.~\ref{Fig02}.

The lowest phonon energy and the axial g-factor of the investigated compound are calculated as $\hbar\omega\approx 1.2$ meV and $g=1.3$, respectively~\cite{Note1}. With $S=15/2$, we have $H_c''\approx0.8$~T, which well agrees with the experimental value (cf. Fig.~\ref{Fig02}). The proposed acceleration mechanism is also supported by other works in that acceleration of the relaxation at similar field strength is prevalent in the magnetic hysteresis measurements of SMMs~\cite{Goodwin2017,Guo2018,Randall2018,Gould2022,Rinehart2011a,Rinehart2011b,Guo2011,Demir2012,Langley2013,Chen2016,Demir2017,Ding2018,Liu2017,Yu2020,Huang2016,Santana2022,Andoni2021,Ding2022,Ling2022}. Moreover, given that the needed data are available, we find that the consistency is quantitative. For example, in Ref.~\cite{Guo2018} the lowest phonon energy is about twice the value here, and the observed $H_c''$ is doubled accordingly.

When computing the relaxation rate, we neglect the nuclear-spin driven quantum tunneling~\cite{Giraud2001,Ishikawa2005,ChenYC2017}, which may cause drop of magnetization around $H=0$ T. When the tunneling is strong, it diminishes the magnetization to nearly zero, and the coercivity is very weak. In this case, the optical-phonon-mediated direct process does not take effect in the stage of magnetization reversal but results in accelerated magnetization toward the saturation (see e.g.~\cite{Huang2016,Santana2022,Andoni2021}).

\paragraph{Roles of intra-molecular exchange interaction.} We focus on a simple setting that the molecule contains two magnetic ions with ferromagnetic interaction. To have a controlled comparison with the single ion case, we assume that the spin-lattice dynamics of the two ions are independent. Namely, $\mathbf{H}_{sp}$ is summation of the respective coupling terms. With this setting, the transition products are null when both $S_{1z}$ and $S_{2z}$ take different values. For example, $\langle S_{1z}=-S,S_{2z}=-S|\mathbf{H}_{sp}|S_{1z}=S,S_{2z}=S\rangle=0$, as the summation form implies that $\mathbf{H}_{sp}$ only acts on one subspace or the other. Transition products $\langle S_{1z}=-S,S_{2z}=S|\mathbf{H}_{sp}|S_{1z}=S,S_{2z}=S\rangle$ and $\langle S_{1z}=S,S_{2z}=-S|\mathbf{H}_{sp}|S_{1z}=S,S_{2z}=S\rangle$ are the counterpart of $\langle S_{z}=-S|\mathbf{H}_{sp}|S_{z}=S\rangle$ in the single ion setting. These transitions constitute significant elements of the transition matrix, and states like $|S_{1z}=S,S_{2z}=-S\rangle$ can be considered as significant intermediate states for the overall magnetic relaxation.

\begin{figure}
	\includegraphics[width=0.75\textwidth]{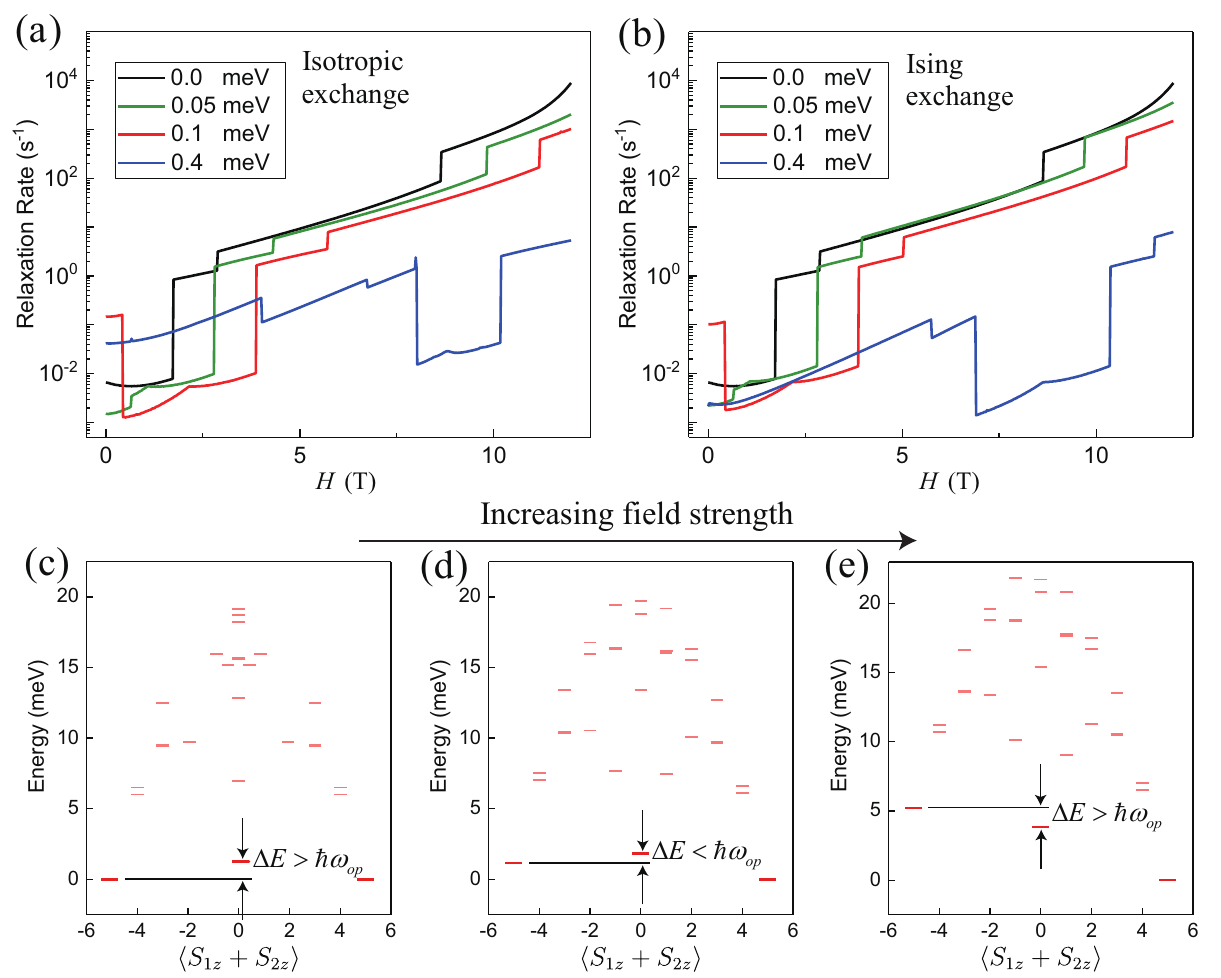}
	\caption{\label{Fig03} (a) Intra-molecular exchange interaction can slow down the relaxation by elevating the intermediate states. When the energy elevation opens up the optical-phonon-mediated direct process, as sketched in (c), significant relaxation enhancement may result. (d) Then, a magnetic field can close the direct process by narrowing the energy gap and results in the drop of the relaxation rate. (e) Further strengthened fields reopen the direct process and causes the leap of the relaxation rate. (b)The Ising exchange has similar effects as the isotropic exchange. The low relaxation rate indicated by the blue curve implies that strong Ising exchange can harness the benefits and largely avoid the harmful effects.}
\end{figure}

The exchange interaction plays two competitive roles. On one hand, it lifts the significant intermediate states, which reduces their thermal accessibility and generally slows down the relaxation. On the other hand, when exchange terms such as $J_xS_{1x}S_{2x}$ and $J_yS_{1y}S_{2y}$ are present, they yield extra state mixing among eigenstates of $S_{1z}$ and $S_{2z}$, besides that from the transverse magnetic anisotropy, since the exchange terms do not commute with $S_{1z}$ and $S_{2z}$. The state mixing may further facilitate relaxation and compromise the coercivity. Accordingly, Ising exchange that has negligible $J_x$ and $J_y$ is beneficial. As shown Fig.~\ref{Fig03}(b), the benefit is obvious when the exchange is strong.

By a mechanism similar to that illustrated in Fig.~\ref{Fig01}(b), elevation of the significant intermediate states shoots up the relaxation rate, when the energy difference with the ground states matches the energy of the significant optical phonons [Fig.~\ref{Fig03}(c)], which activates the optical-phonon-mediated direct process. Then, performing a magnetic field and tuning its strength can reduce the energy difference, break the energy match [Fig.~\ref{Fig03}(d)] and turn off the rapid relaxation, resulting in drop of relaxation rate. A regime of low relaxation rate follows, until the direct process is opened up again by further strengthened fields [Fig.~\ref{Fig03}(e)]. When the exchange is strong, although the undesirable direct process is always on before a high field turns it off, the relaxation is kept relatively slow, because the high-lifted significant intermediate states imply low availability of the associated phonons.

We note that this setting is barely explored in previous experiments, since the anions bridging the exchange usually had unpaired electrons that lead to direct exchange interaction with the metallic ions~\cite{Rinehart2011a,Rinehart2011b,Guo2011,Demir2012,Langley2013,Demir2017}. Those are mixed-valence compounds investigated in the following. In the setting of double ionic moments, one way to enhance the coercivity is to mildly lifted the 
significant intermediate states with a relatively weak field. The optimal energy gap with the ground states is $\Delta E\lesssim\hbar\omega_{op}$, i.e., smaller than $\hbar\omega_{op}$ but close to it. Then, the optical-phonon-mediated direct process is initially off and activated when reaching the configuration in Fig.~\ref{Fig03}(e) with $H''_c\approx\hbar\omega_{op}/S\mu_B g$. The critical field strength is doubled compared to the single ion setting. We can also enhance the coercivity by highly elevating the significant intermediate states with strong Ising exchange [blue curve in Fig.~\ref{Fig03}(b)], which could result in a very strong coercivity.

\paragraph{Magnetization reversal in mixed-valence compounds.} In regard to the molecules studied in Ref.~\cite{Gould2022} we consider a structure where two ionic magnetic moments are coupled to a bonding electron through ferromagnetic exchange interaction~\footnote{In Sec. II C of the Supplementary Materials we argue why the case of antiferromagnetic exchange is generally similar.}. As there is no exchange between the ionic moments, the significant intermediate states consist mainly of opposite ionic moments are not lifted effectively, so the consequential reduction of relaxation rate is weak. On the other hand, the state mixing due to the isotropic exchange lead to sizable relaxation rate increase~\cite{Note1}. In contrast, Ising exchange can maintain low relaxation rate and enhance $H_c''$ by elevating the intermediate state~\cite{Note1}. This explains why magnetic hysteresis was usually observed in mixed-valence compounds with Ising exchange ~\cite{Rinehart2011a,Rinehart2011b,Guo2011,Demir2012,Langley2013,Demir2017}. The remarkable coercivity enhancement in Ref.~\cite{Gould2022} should arise from other factors such as the doubled magnetic anisotropy.

\begin{figure}
	\includegraphics[width=0.6\textwidth]{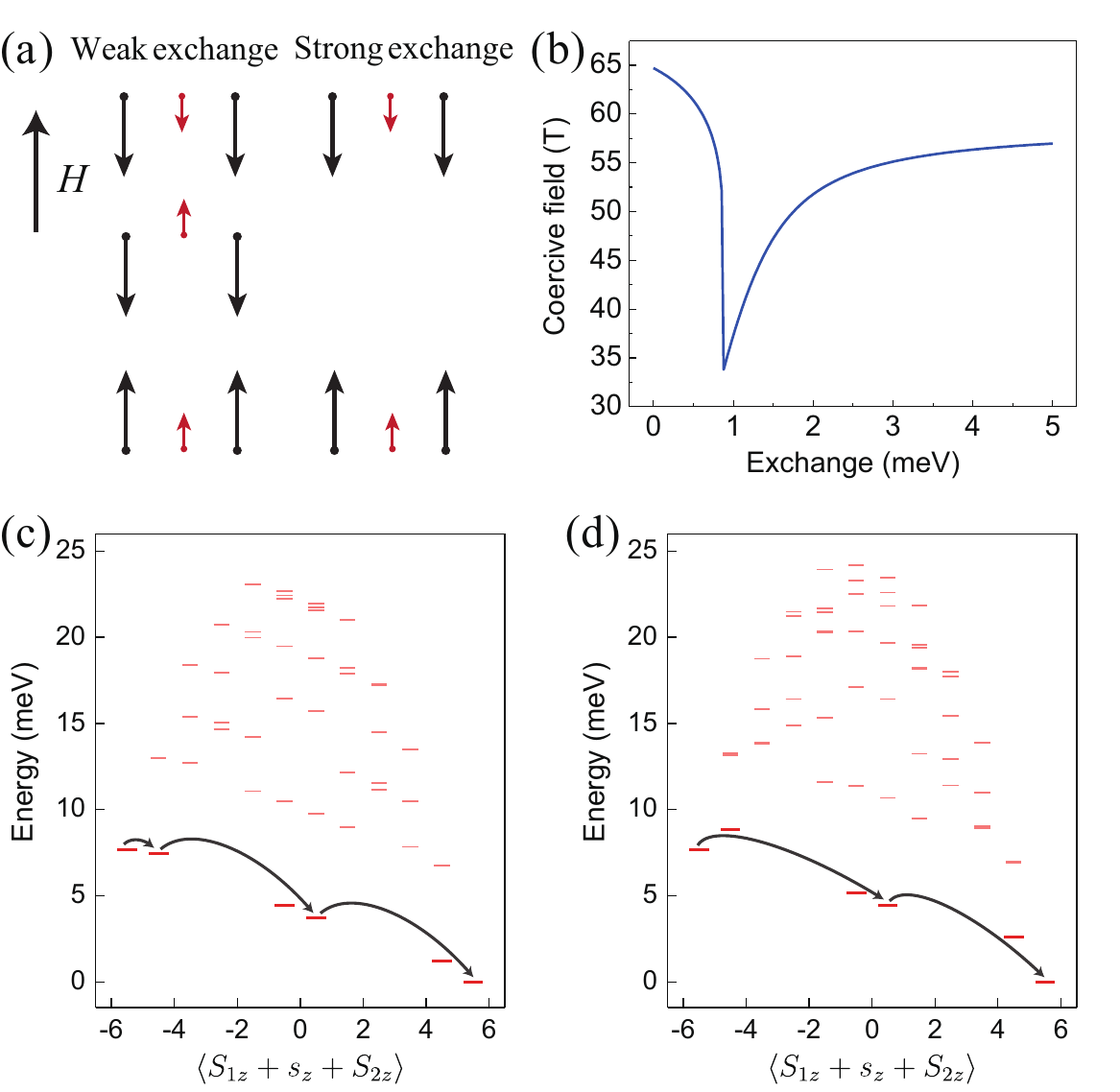}
	\caption{\label{Fig04} (a) Schematics of magnetization reversal processes in the cases of weak and strong exchange, respectively. (b) By the classical approach the transition between the two phases yields a turning point of coercivity. (d) The quantum dynamical simulation suggests that strong exchange does suppress the intermediate process representing the pre-spin-flip, which takes effect in the weak exchange case, as indicated by the left arrow in (c).}
\end{figure}

The magnetic moment of the binding electron induces a peculiar magnetization reversal process. In the classical picture, when the exchange is weak, one can expect that the spin of the bonding electron would be flipped by a relatively weak magnetic field before the two ionic magnetic moments are reversed by a stronger field [Fig.~\ref{Fig04}(a)]. This is because the reversal of the ionic magnetic moments requires overcoming the anisotropy barrier, but the electron spin is stabilized solely by the exchange interaction. Consequentially, the pre-spin-flip of the bonding electron would facilitate the overall magnetization reversal, since the ferromagnetic exchange implies that the flipped electron spin counters part of the anisotropy barrier. When the exchange is strong enough, the pre-spin-flip is suppressed, and the magnetic moments are reversed in whole. Fig.~\ref{Fig04}(b) presents coercivity by the classical approach, which manifests the transition point between these two phases.

Interestingly, this classical picture has quantum correspondence in terms of magnetic relaxation. As shown in Fig.~\ref{Fig04}(c) and Fig.~\ref{Fig04}(d), the key lies in the ground state built on $|S_z=-2S-1/2\rangle$ (the first from the left that has been lifted by the magnetic field), and the excited state mainly composed of $|S_z=-2S+1/2\rangle$ (the second from the left). Here, $S_z$ is the total $z$ component, $S_z=S_{1z}+s_z+S_{1z}$. Inspection of the wavefunctions indicates that the difference in $S_z$ does arise from spin flip of the bonding electron~\cite{Note1}. When the exchange is weak, the energy gap between these two states is small, magnetic field could efficiently reduce the gap and even make the energy of $|S_z=-2S-1/2\rangle$ higher [Fig.~\ref{Fig04}(c)]. In this case, our simulations~\cite{Note1} shows that the system fleets from $|S_z=-2S-1/2\rangle$ to $|S_z=-2S+1/2\rangle$ in the beginning stage of the relaxation [Fig.~\ref{Fig04}(c)], representing flip of the electron spin. When the exchange becomes stronger, $|S_z=-2S+1/2\rangle$ is less accessible from $|S_z=-2S-1/2\rangle$ because of the larger energy gap. Then, the relaxation does not undergo this intermediate state [Fig.~\ref{Fig04}(d)].

\paragraph{Conclusions and outlook.} We have showed that the magnetic field can substantially modulate the magnetic relaxation of SMMs. Because of the prominent role of optical phonons, activation and suppression of related relaxation pathways due to energy match and mismatch should be one of our focuses for understanding and tuning the coercivity. Especially, in SSMs with slow magnetic relaxation, the tuning on of direct process mediated by optical phonons is the reason for the accelerated relaxation that determines the measured coercivity. The level-crossing-induced relaxation rate skyrocketing is theoretically sound and can be observed with proper experimental setup. While it can be easily obscured in measurement with scanning fields and does not actually concern the coercivity of SMM with strong zero field splitting, it offers a means of ultra-fast magnetization reversal of SMMs.

A classical view~\cite{Brown1945,Stoner1991} of demagnetization at macroscopic scale is coherent evolution of magnetic moments where each needs to overcome the magnetic anisotropy barrier. A core lesson from our results is that spin-phonon relaxation involves shortcut channels that can be significantly enhanced by magnetic fields. An implicit idea about magnetic relaxation of SMMs is that the magnetic moment(s) in each unit cell evolves incoherently, which seems inapplicable to systems with long range magnetic orders. Nevertheless, transition among spin states of local magnetic moment(s) may compromise the long-range order in the first place and hence reduce the coercivity. Concrete investigation of this intuition may require developments of theoretical tools beyond the framework of magnon-phonon interaction that presumes long range magnetic order and coherent interactions.

\begin{acknowledgments}
\end{acknowledgments}

\bibliography{coercivity}

\end{document}